\documentclass[10pt,preprint]{aastex}
\def\poin{Poincar\'e}

\def\a0{a_0}

\def\beq{\begin{equation}}
\def\eeqno#1{\label{#1}\end{equation}}

\def\az{a_{0}}

\def\l0{\ell_{0}}

\def\rar{\rightarrow}

\def\r{\rho}

\def\S{\Sigma}
\def\drt{d^3r}
\def\f{\phi}
\def\grad{\vec\nabla}
\def\div{\vec \nabla\cdot}
\def\gf{\grad\phi}

\def\vr{\textbf{r}}

\def\vv{\textbf{v}}
\def\vp{\textbf{p}}
\def\vP{\textbf{P}}
\def\vV{\textbf{V}}
\def\va{\textbf{a}}

\def\vgz{\textbf{g}_0}

\def\vR{\textbf{R}}
\def\gz{g_0}

\def\cm{{~\rm cm}}

\def\l{\lambda}
\def\L{\Lambda}

\def\a{\alpha}
\def\b{\beta}

\def\d{\delta}
\def\m{\mu}
\def\n{\nu}
\def\f{\phi}

\def\az{a_{0}}
\def\baz{\bar a_{0}}
\def\haz{\hat a_{0}}

\def\vP{{\bf P}}

\def\rar{\rightarrow}
\def\dSST{$dS^4$}

\usepackage{natbib}
\usepackage{amssymb}
\usepackage{epsfig}
\begin{document}

\title{The MOND limit from space-time scale invariance}

\author{Mordehai Milgrom}
\affil{Center for Astrophysics, Weizmann Institute, Rehovot 76100,
Israel}
\begin{abstract}{The MOND limit
is shown to follow from a requirement of space-time scale invariance
of the equations of motion for nonrelativistic, purely gravitational
systems; i.e., invariance of the equations of motion under
$(t,\vr)\rar(\l t,\l\vr)$ in the limit $\az\rar\infty$ . It is
suggested that this should replace the definition of the MOND limit
based on the low-acceleration behavior of a Newtonian-MOND
interpolating function. In this way, the salient, deep-MOND
results--asymptotically flat rotation curves, the
mass-rotational-speed relation (baryonic Tully-Fisher relation), the
Faber-Jackson relation, etc.--follow from a symmetry principle. For
example, asymptotic flatness of rotation curves reflects the fact
that radii change under scaling, while velocities do not. I then
comment on the interpretation of the deep-MOND limit as one of
``zero mass'': Rest masses, whose presence obstructs scaling
symmetry, become negligible compared to the ``phantom'', dynamical
masses--those that some would attribute to dark matter. Unlike the
former masses, the latter transform in a way that is consistent with
the symmetry. Finally, I discuss the putative MOND-cosmology
connection in light of another, previously known symmetry of the
deep-MOND limit. In particular, it is suggested that MOND is related
to the asymptotic de Sitter geometry of our universe. It is
conjectured, for example, that in an exact de Sitter cosmos,
deep-MOND physics would exactly apply to local systems. I also point
out, in this connection, the possible relevance of a de
Sitter-conformal-field-theory (dS/CFT) duality.}
\end{abstract}

\section{Introduction}
\label{section1}
 MOND has been advanced as an alternative paradigm
to Newtonian dynamics, whose original motivation was to explain the
mass discrepancies in galactic systems without invoking dark matter
(Milgrom 1983, for reviews see Sanders and McGaugh 2002, Bekenstein
2006, and Milgrom 2008). The paradigm is constructed on three
premises: (i) There appears in non-relativistic physics a new
constant, $\az$, with the dimensions of acceleration. (ii) A
correspondence principle is required that guarantees restoration of
Newtonian physics when we formally take $\az\rar 0$ in all the
equations of motion (just as relativistic physics tends to classical
physics when we formally take the speed of light $c\rar \infty$).
This matches the original introduction of MOND as a modification of
Newtonian dynamics only when accelerations in the system are not
much larger than $\az$. (iii) Some constraints on the behavior of
the theory in the opposite, deep-MOND limit. This limit is achieved
by formally taking $\az\rar \infty$ in the equations of motion.

Initially (Milgrom 1983 and onward) this deep-MOND limit was defined
in terms of some MOND interpolating function of the acceleration,
call it $\m(a/\az)$, which is required to have the limit
$\m(x)\approx x$ for $x\ll 1$, corresponding to $\az\rar \infty$.
This requirement is based solely on the axiom of MOND that isolated
masses should have an asymptotically flat rotation curve, and so the
interpolating function in question has to be related to rotation
curves. Some effective MOND formulations (such as that of Bekenstein
and Milgrom 1984) may indeed be defined by just one interpolating
function. But in general the theory should not be defined by an
interpolating function: The theory of relativity and quantum
mechanics, which also constitute departures from classical physics,
do not hinge on a fundamental function that connects them smoothly
to the classical regime. To be sure, interpolating functions have to
appear in specific contexts of MOND just as they do in relativistic
or quantum expressions\footnote{Examples in relativity are the
Lorentz factor, and various functions describing the behavior of
particles near black holes, which tend to the classical form when
$c\rar\infty$. In the quantum-classical case, the black-body
function, the expression for the specific heat of solids, or barrier
transmission probabilities by tunneling are examples of
interpolating functions that tend to the classical expression for
$\hbar\rar 0$.}; but, these are not necessarily fundamental and may
even have different asymptotic behaviors.
\par
Clearly, it is desirable to define the MOND limit in a general way,
without recourse to interpolating functions. Starting in Milgrom
(2001), and more elaborately in Milgrom (2008), I defined
the MOND limit for purely gravitational systems by the
requirement that for $\az\rar \infty$, the equations of motion are
writable in a form where the constants $\az$, $G$, and all masses in
the problem, $m_i$, appear only in the products $m_iG\az$. I also
demonstrated there how this dictate, with the other two tenets of
MOND, leads to many of the salient MOND predictions. While this
requirement is operationally sound, it may appear artificial [see
discussion around eq.(\ref{iigu}) below].
\par
Here I point out that the third MOND tenet, defining the deep-MOND
limit, can be based on a symmetry requirement: A (nonrelativistic) MOND theory for a purely
gravitational system has to becomes space-time scale invariant in the
limit $\az\rar\infty$; i.e., invariant under $(t,\vr)\rar(\l
t,\l\vr)$ (leaving intact dimensioned constants of the system such
as masses, $G$, and $\az$--scaling is not a mere change of units).
\par
This belated observation is important because (i) it may help put
MOND on more sound footings, showing that it need not be imposed as
an ad-hoc dictate of phenomenology; (ii) identifying an underlying
symmetry may help extend MOND to non gravitational systems for which
we have no phenomenological guidance, at present; (iii) identifying
a partial symmetry valid in the deep-MOND regime may lead us to
identify a larger symmetry in this regime, or to identifying a
symmetry group for the full MOND theory, and this will help
constrain the MOND theory itself; (iv) this description is more
readily amenable to direct predictions.

There are several, possibly unrelated, facts that hint to the
possibility that MOND is underlaid by a new symmetry, so that even
as a non-relativistic theory it enjoys a space-time symmetry other
than the Galilei group. For example, I showed in Milgrom (1994) that
a MOND theory that is derived from an action and has the Newtonian
and the deep-MOND limits as required, and enjoys Galilean
invariance, cannot be local. So perhaps a local theory is possible
with an underlying symmetry other than Galilei. In Milgrom (1997) I
noted that the deep-MOND limit of the equation for the gravitational
potential, in the formulation of Bekenstein and Milgrom (1984), is
invariant to conformal space transformations. In Milgrom (2005) I
commented on the superficial similarity between the MOND kinematics
revolving around accelerations with the special relativistic
kinematics revolving around velocities, with $\az^2$ playing the
role of $-c^2$, and with the new underlying symmetry having perhaps
to do with de Sitter (hereafter dS) space-time\footnote{Even if the
``dark energy'' is a ``cosmological constant'', is not the sole
contribution to the matter content of our universe. We thus do not
live in an exact de Sitter space-time, and this might result in the
breaking of the symmetries relevant to MOND.} in which $\az$ is a
proxy for $c^2\L^{1/2}$.
\par
On the other hand, the symmetry discussed here may be incidental,
and specific to the purely gravitational case. This would be similar
to the equations of motion of Newtonian gravity being invariant
under $(t,\vr)\rar (\l t,\l^{2/3}\vr)$ (more on this in the
discussion section), or to the problem involving harmonic forces
being invariant under $(t,\vr)\rar (t,\l\vr)$. In these two cases
the action governing the theories is not invariant, but multiplied
by a constant factor, under the corresponding scaling. This means
that while the equations of motion are invariant, there are no
conservation laws associated with the symmetry.  If the scale
invariance described here for the deep-MOND limit is of the same
kind, and does not have fundamental underpinnings, it is still a
useful tool, especially since MOND is inherently nonlinear, and is,
otherwise, hardly amenable to deduction of exact results.

In section \ref{sect:scaling} I explain how the assumed space-time
scaling invariance leads to the earlier formulations of the third
tenet of MOND. In section \ref{sect:phenomen} I show how scaling directly
begets a variety of predictions. Section \ref{sect:zromass} describes the
deep-MOND limit as a formal zero mass limit. In section \ref{sect:cosmo} I discuss
possible connections of MOND with cosmology and the interpretation
of the deep-MOND limit. Section \ref{sect:disc} is a discussion.

\section{Space-time scale invariance and the deep-MOND limit}
\label{sect:scaling} Consider a purely gravitational system in the
deep-MOND regime. For convenience, I assume that the system is made
of discrete masses $m_i$\footnote{For a truly point mass, the field
near the mass is not in the deep MOND regime; so, when we speak of
masses we have to view them as larger than their MOND transition
radius $r_t\equiv (MGg/\az)^{1/2}$, or consider only the regions
outside $r_t$ for all masses. For reference, note that the
transition radius of a proton is $\sim10^{-12}\cm$.}. The assumption
of pure gravity implies that the only physical constants that appear
in the description of the system are $G$, $\az$, and masses $m_i$. I
do allow forces that constrain some of the masses to move on
prescribed trajectories (such as keeping them at fixed positions) as
such forces do not introduce additional dimensioned constants (the
prescribed trajectories will always undergo scaling along with the
dynamical ones). The problem we study consists of determining the
trajectories of the gravitating particles, given some initial
conditions. These are determined from equations of motion of the
general form (combining dynamics with gravity)

\beq  \mathcal{F}_k[m_i,G,\az,\vr_i(t)]=0,~~~~k=1,2,3,...~~.\eeqno{ii}

Here, $\vr_i(t)$ stand for trajectories of the particles, including
those prescribed, and $\mathcal{F}_k$ are general functionals of the
trajectories, so they can contain any number of derivatives of
$\vr_i(t)$,  or be non-local functionals of them. The statement of
the symmetry required in the deep-MOND limit is then: In the limit
$\az\rar \infty$, if $\vr_i(t)$ constitute an allowed configuration
of the particle motions, then so do
 $\hat\vr_i(\hat t)=\l\vr_i(t)=\l\vr_i(\hat t/\l)$ (corresponding to  $\hat\vr_i=\l\vr_i$, and $\hat t=\l
 t$), with the appropriate scaling of the initial conditions. In other
words, if eq.(\ref{ii}) holds for some $\vr_i(t)$ then it is also
true that

\beq
\mathcal{F}_k[m_i,G,\az,\l\vr_i(t/\l)]=0,~~~~k=1,2,3,...~~.\eeqno{iii}
Note that the constants and the masses remain
intact\footnote{Sometimes we describe part of the system as a fluid,
characterized by a density and velocity fields, $\r(\vr,t)$ and
$\vv(\vr,t)$, respectively (instead of by masses and their
positions). These fields then transform under scaling as
$\r(\vr,t)\rar \l^{-3}\r(\vr/\l,t/\l)$ and
$\vv(\vr,t)\rar\vv(\vr/\l,t/\l)$}.

 The equations of motion
may also involve gravitational fields that mediate gravity, with
their own transformation properties under scaling that insure the
invariance of  the equations of motion. But, I concentrate on the
observable motion of the masses $\vr(t)$, and on what the symmetry
implies for them. (One may think, for example, of equations of
motion from which the gravitational fields have been eliminated by
expressing them in terms of the particle trajectories and
substituting them back in the equations of motion.)
\par
To make contact with the earlier definition of the deep-MOND limit,
note that from dimensional considerations alone, we are free to
multiply all quantities with dimensions of mass, lengths, and time
by arbitrary, possibly dimensioned, constants, $\a,~\eta,$ and
$\xi$, without impairing the validity of an equation. Equation
(\ref{ii}) can thus be written as

\beq  \mathcal{F}_k[\a
m_i,\eta^3\xi^{-2}\a^{-1}G,\eta\xi^{-2}\az,\eta\vr_i(t/\xi)]=0.\eeqno{iiifu}
Take now $\xi=\eta=\az$ and $\a=G\az$, and impose the symmetry to
get

\beq  \mathcal{F}_k[(m_iG\az,1,1,\vr_i(t)]=0,\eeqno{v} which proves
that the equations governing the motion of particles can be brought
to a form in which $m_i,~G,~\az$ appear only in the products
$m_iG\az$. This is the formulation of the third MOND tenet as given
in Milgrom (2008), which is now seen to follow from scale
invariance\footnote{Note that this conclusion does not change if the
theory also involves physical constants with the dimensions of
velocity (such as the speed of light), whose values do not change
under scaling. But our subsequent results do assume the irrelevance
of such additional constants.}. Clearly the opposite is also true,
as stated in Milgrom (2008). The above result simply reflects the
fact that any combination of $M$, $G$, and $\az$ whose value is
invariant under joint scaling of the length and time units must be a
function of $MG\az$.
\par
As an example, the very first description of the deep-MOND limit
(Milgrom 1983) giving the acceleration of a test particle at a
distance $R$ from a point mass $M$ as:
 \beq (a/\az)\va=-MG\vR/R^3  \eeqno{jula}
is clearly scale invariant in the above sense.
\par
So far I have only discussed the equations of motion, which are more
directly related to phenomenology. What can be said about the effect
of scaling on an underlying action?
 To obtain scale invariance of
the equations of motion it is sufficient that the action is
multiplied by a constant under scaling. However, if the action is
not truly invariant, only multiplied by a constant, the symmetry
does not imply a conservation law. We don't know what action will
govern the final MOND theory; but start, as an example, with the
Newtonian action governing a nonrelativistic system of gravitating
point masses:

\beq S=-{1\over 8\pi G}\int dt \drt~(\gf)^2- \sum_i m_i\int
dt\f[\vr_i(t)] + \sum_i m_i\int dt ~v_i^2(t)/2. \eeqno{action}
 Its different terms cannot transform in the same way under scaling:
Comparing the last two terms, we see that for them to transform in
the same way, $\f$ would have to have zero dimension under scaling
(i.e., transform as $\f(\vr,t)\rar \f(\vr/\l,t/\l)$; but, then, the
first term does not transform like the other two.
\par
One way to effect uniform transformation properties for the three
terms, with the help of $\az$, is to modify the first
term\footnote{I call such modifications ``modified gravity'' because
they modify the gravitational field, leaving Newton's second law
intact.} replacing $(\gf)^2$ by another Lagrangian density that has
three powers of length in the denominator\footnote{In our
nonrelativistic case, time derivatives of the potential do not
appear, as the potential adjusts itself instantaneously to motion of
the masses; so, the field equations hold time by time.}. For example
it can be proportional to $|\gf|^3$, or if we allow higher
derivatives of the potential, $(\Delta\f)^{3/2},$
 $(\f_{,i,j}\f_{,k}\f_{,l} \d^{ik}\d^{jl})^{3/4},$ etc.. However, since
we have only $\az$ at our disposal as an additional dimensioned
constant we can only form modified Lagrangian densities involving
the first derivatives of $\f$ (if we allow only derivatives of $\f$,
not $\f$ itself, to appear). Our assumed scale invariance thus leads
to a unique Lagrangian for the deep-MOND limit in this single
potential case; i.e., $|\gf|^3/\az$, which is indeed the deep-MOND
limit of the Bekenstein and Milgrom (1984) formulation. The field
equation for the gravitational potential then becomes a modification
of the Poisson equation with the Laplacian replaced by the
3-Laplacian $\div(|\gf|\gf)$.

Interestingly, I showed in Milgrom (1997) that the deep-MOND limit
of the Bekenstein and Milgrom (1984) formulation for the field
equation for the potential is, in fact, invariant under the whole
group of conformal {\it space} transformation (which includes space
scaling)\footnote{This turns out to be the case in all the above
mentioned scale invariant gravitational Lagrangian containing higher
derivatives of the potential, not only the MOND case. It is well
known that scale invariance, together with rotational invariance,
implies conformal invariance in a large class of theories. Our
theory is not invariant to rotations in space-time--i.e., is not
Lorentz invariant; so, the space-time scale invariance does not
imply full space-time conformal invariance, but perhaps a
generalization of it will.}. The above deep-MOND action is not
invariant, but is multiplied by a constant under scaling.
\par
Another option to construct a MOND action is to  modify the kinetic
action of particles--so called ``modified inertia'' (Milgrom 1994):
Replace the last term in expression (\ref{action}) by a kinetic
action of the form
 \beq S_k=M\tilde S_k[\az, q_1,q_2,..., \vr_1(t),\vr_2(t),...],  \eeqno{kinet}
where $M=\sum_i m_i$, and, on dimensional grounds, $\tilde S_k$,
which has dimensions of length$^2$/time, can only be a function of
the mass ratios $q_i=m_i/M$, in addition to its being a function of
$\az$ and a functional of the particle trajectories $\vr_i(t)$. In
order to insure universality of free fall we require that $\tilde
S_k$ is symmetric under interchange of particle indices (a possible
form is $\tilde S_k=\sum_i q_i \tilde s_k[\az,\vr_i(t)]$). The first
two terms in expression (\ref{action}) have the same transformation
properties if $\f$ is given dimension $-1$ under scaling [by decree,
complementing the transformation rule of $t$ and $\vr$ by
$\f(\vr,t)\rar \l^{-1}\f(\vr/\l,t/\l)$]. Then, to have the same
transformation properties as the first two we have to have in the
limit $\az\rar\infty$,
 \beq \tilde S_k\rar \az^{-1} \sigma[q_1,q_2,..., \vr_1(t),\vr_2(t),...],  \eeqno{kinetop}
where $\sigma$ is a functional of the trajectories (and function of
the mass ratios) with dimensions of length$^3$/time$^3$. Multiplying
the action in eq.(\ref{action}) by $G\az^2$ and absorbing one power
of $\az$ in the definition of $\f$, defining $\psi=\az\f$,  we get,
in the limit, an action in which only $m_i G\az$ appear:
 \beq S=-{1\over 8\pi }\int dt
\drt~(\grad\psi)^2- \sum_i m_iG\az\int dt\psi[\vr_i(t)] +
MG\az\sigma[q_1,q_2,..., \vr_1(t),\vr_2(t),...]. \eeqno{actionv}
\par
 As in the Newtonian case,
the potential can be eliminated from the action: First note that for
solutions of the equations of motion the first term in the action is
$-1/2$ times the second, which, in turn, can be explicitly expressed
in terms of the positions to give
 \beq S=-\int dt \Phi[\vr_i(t)] +
MG\az\sigma, \eeqno{actionvi} where the potential energy
 \beq \Phi=-\sum_{i<j}{m_im_j(G\az)^2\over |\vr_i(t)-\vr_j(t)|}.
 \eeqno{mijta}

This ``modified-inertia'', deep-MOND action in expressions
(\ref{actionv}) or (\ref{actionvi}) is truly invariant under scaling
$(t,\vr,\psi)\rar(\l t,\l\vr,\l^{-1}\psi)$. For reasons explained in
Milgrom (1994) we ought to concentrate on nonlocal forms of
$\sigma$, in which case no conservation law is implied. For
completeness I mention that for local kinetic actions, the symmetry
would imply a conservation law. For example, if the kinetic action
can be written in terms of a kinetic Lagrangian depending on
derivatives up to order $n$:
 \beq MG\az\sigma=\sum_i \int dt L[\dot \vr_i,\ddot\vr_i,...,\vr_i^{(n)}], \eeqno{lagra}
the conserved quantity can be shown to be
 \beq Q\equiv -Ht+\sum _i\sum_{m=1}^n
\vr_i^{(m-1)}\cdot \vp_i^m. \eeqno{conser} Here, the energy $H$ and
the momenta $\vp_i^m$ are those introduced by Ostrogradski,
expressions for which are given, e.g., in Milgrom (1994).
(Invariance to time translations implies the conservation of $H$,
and to space translations the conservation of $\vP^1=\sum_i
\vp_i^1$.)
\par
The division into ``modified gravity'' and ``modified inertia'' is
not exhaustive, and, not even always well defined, as we can devise
simultaneous modifications of all terms in the action. Perhaps a
more appropriate distinction is between modifications that are
``modified gravity'' and those that are not. In the former,
nonrelativistic gravity is still described by a potential, $\phi$,
in which all masses still follow $\va=-\grad\phi$, but the potential
is determined by an equation other than Poisson's. In the latter class
of theories this is not so, and the above ``modified inertia''
action is an example of such a theory.
\par
For the relativistic case we might call ``modified gravity'' a
theory that can be cast as a metric one, with the standard coupling
of matter degrees of freedom to the metric (implying geodesic
motion). It departs from General Relativity in that the metric is
not determined from the Einstein equations. The nonrelativistic
limit of such theories is then a ``modified gravity'' theory in the
above sense.

\section{Predicted phenomenology}
\label{sect:phenomen}
 \subsection{Asymptotic rotation speeds}
\label{subsec1}
  There are several important
implications for gravitational systems in the deep-MOND regime that
can be deduced directly from the assumed scaling invariance. In the
first place, it tells us that if $\vr(t)$ is a trajectory of a point
body in a configuration of point masses\footnote{Under scaling, an
extended mass changes its size and density such that the total mass
remains the same. A point mass remains a point mass of the same
value.} $m_i$ at positions $\vr_i(t)$ (which can be taken as fixed,
for example), then $\hat\vr(t)=\lambda\vr(t/\lambda)$ is a
trajectory for the configuration where $m_i$ are at
$\lambda\vr_i(t/\lambda)$, and the velocities on that trajectory are
$\hat\vV(t)=\vV(t/\lambda)$.  It follows from this, for example,
that the rotational speed in an orbit around an isolated mass
becomes independent of the size of the orbit in the large size
limit: Under scaling the orbit changes its size, but the velocities
do not change. The extent of the attracting mass also scales, but
for large orbital radii, the size of the central mass becomes
immaterial and does not affect the motion\footnote{I assume that the
effect of a mass becomes independent of its internal structure,
including its extent, when this extent becomes very small. Otherwise
the concept of point mass loses its meaning anyway.}. This leads to
the asymptotic flatness of rotation curves of isolated galaxies. In
the present formulation of the MOND limit, this cornerstone axiom of
MOND is thus simply a reflection of the fact that velocities do not
change under space-time scaling, while distances do.

Accelerations scale as $\l^{-1}$; so, in the deep-MOND limit, when a
physical orbit is scaled up to another the accelerations scale as
the inverse of the orbit's size. This scaling of the acceleration
does not, in general, relate accelerations at different point on the
same orbit. However, in the class of theories constituting
``modified gravity'', namely, theories in which the gravitational
field is modified, and the acceleration depends only on position,
the above scaling does tell us that the acceleration of a test mass
in the field of single point mass decreases as the inverse of the
distance from that mass\footnote{This also follows from the
statement, made above, that in modified gravity the potential has to
have zero scaling dimension; this in turn, implies that it has to be
logarithmic in the distance in the deep-MOND limit.}. It also tells
us, more generally, that the gravitational force between two point
masses must decrease as the inverse of the distance between them.

 \subsection{Mass-velocity relations}
\label{subsec2}
 A change of the time units by a factor $\xi$ in
eq.(\ref{v}) gives
 \beq
\mathcal{F}_k[(\xi^{-4}m_iG\az,1,1,\vr_i(t/\xi)]=0,\eeqno{vfu} which
tells us how orbital characteristics change with scaling of the
masses.
\par
Before discussing the implications of this relation I derive it in
another way, which gives an added insight. The appearance of
dimensioned constants in physics is, many times, artificial, and
results from our insistence on inventing new units for quantities
that can be measured in existing ones. But such constants may,
actually, be useful as bookkeeping tools, as in the examples above,
and also in helping us implement certain limits such as the
nonrelativistic limit, or, as here, the MOND limit. Alternatively,
we may work in units in which $\az=G=1$ are
dimensionless\footnote{Defining the MOND and Newtonian limits then
becomes somewhat more elaborate: we have to identify all quantities
with the role of acceleration, and require that they are much
smaller than 1, or much larger than 1, respectively.}. Then, lengths
have dimensions of time squared, and masses have dimensions of time
to the fourth power\footnote{Under change of the units of time we
then have: $t\rar\xi^{-1} t,~\vr\rar\xi^{-2}\vr,~m\rar\xi^{-4}$; but
under scaling we still have $t\rar \l t,~\vr\rar\l\vr,~m\rar m$ .}.
The MOND equations of motion (\ref{ii}) are now of the form
 \beq
\hat\mathcal{F}_k[m_i,\vr_i(t)]=0,~~~~k=1,2,3,...~~,\eeqno{iigu}
where we cannot even describe the MOND limit in terms of how the
constants appear. Scale invariance can be implemented as follows:
under change the units of time by a dimensionless factor $\xi$,
eq.(\ref{iigu}) becomes \beq \hat\mathcal{F}_k[\xi^{-4}
m_i,\xi^{-2}\vr_i(\xi t)]=0,~~~~k=1,2,3,...~~.\eeqno{iigul}
 Now apply the scale invariance (valid in the deep-MOND limit), which tells us that
 $\vr_i(\xi t)$ in eq.(\ref{iigul}) can be replaced by
 $\l\vr_i(\xi t/\l)$. Choosing $\l=\xi^2$
gives \beq\hat\mathcal{F}_k[\xi^{-4}
m_i,\vr_i(t/\xi)]=0,~~~~k=1,2,3,...~~,\eeqno{iiigul} which has the
same content as eq.(\ref{vfu}). Thus, if we have a solution
$\vr_i(t)$ for a given choice of the masses, then $\vr_i(t/\xi)$ is
a solution for a system where all the masses are multiplied by
$\xi^{-4}$. The orbits remain the same in space but the bodies
traverse them in times multiplied by a factor $\xi$, speeds that are
multiplied by $\xi^{-1}$, and accelerations that are multiplied by
$\xi^{-2}$ (for the appropriately scaled initial conditions). More
generally, choosing $\l=\xi^\a$ gives \beq\hat\mathcal{F}_k[\xi^{-4}
m_i,\xi^{\a-2}\vr_i(t\xi^{1-\a})]=0,~~~~k=1,2,3,...~~.\eeqno{iiiguli}
The velocities for all these solutions still scale as the fourth
root of the masses. If the theory is such that the trajectory
$\vr(t)$ of a test particle does not depend on its mass--as required
by the universality of free fall, and as I assume all along--we do
not have to scale the mass of test particles.
\par
From this scaling with mass follows the MOND mass-asymptotic-speed
relation (underlying the baryonic Tully-Fisher relation)
 \beq V^4_{\infty}=MG\az. \eeqno{tf}
This corollary also tells us that a MOND theory must be nonlinear:
Since accelerations here scale as the square root of the mass
(unlike the Newtonian scaling with mass) the acceleration, light
bending, etc., that are produced by several masses is not the sum of
those produced by the individual masses.
 \subsection{Virial relations}
\label{subsec3}
 This scaling of velocities with mass also tightly
constrains the form of the deep-MOND virial relation: Consider some
measure of the mean squared velocity, $\langle V^2\rangle$, in a
stationary, many-body, self-gravitating, bound system of total mass
$M$. For a system deep in the MOND regime the above scaling tells us
that
 \beq \langle
V^2\rangle^2=MG\az Q, \eeqno{mila} where $Q$ is a function of
dimensionless attributes of the system (such as mass ratios,
geometrical factors, velocity anisotropies, etc.), which, in
particular, does not depend on the overall size of the system. The
function $Q$ also depends on the particular formulation of MOND at
hand. For example, for the formulation of Bekenstein and Milgrom
(1984), the space, conformal invariance of the deep-MOND limit was
shown (Milgrom 1997) to lead to the following general, exact result:
If $\langle V^2\rangle$ is the mass weighted, mean square velocity
$\langle V^2\rangle=\sum m_iV^2_i/M$, then

 \beq Q={4\over
9}(1-\sum_i q_i^{3/2})^2 , \eeqno{opta} which depends only on mass
ratios in the system, $q_i$, and becomes a number, $Q=4/9$, when all
masses are small compared with the total mass
[$Q=(4/9)(1-N^{-1/2})^2$ for a system of $N$ equal mass bodies].
This scaling law also underlies the Faber-Jackson relation for low
acceleration systems. As explained in Milgrom (2008), with the help
of the other MOND tenets, this extends to a predicted Faber-Jackson
correlation for spherical systems in general, provided they can be
approximately described as isothermal spheres.

I note in passing that the above virial relation as applied to thin
disc galaxies in the deep-MOND regime reads

\beq \langle V^2\rangle^2 ={4\over 9}MG\az, \eeqno{jipa} where
$\langle V^2\rangle \equiv \int_0^{\infty}2\pi r\Sigma(r)V^2(r)dr$,
with $\Sigma(r)$ and $V(r)$ are, respectively, the normalized mass
surface density ($2\pi \int r\Sigma(r)dr=1$), and the rotational
speed in the disk. This resembles the mass velocity relation
(\ref{tf}), but, in fact, it is a rather different, deep-MOND result
that follows from the same scaling property: it involves the rms
velocity, not the asymptotic one, it is specific to the Bekenstein
\& Milgrom (1984) formulation, and it applies only to deep-MOND
galaxies unlike relation (\ref{tf}), which is universal.

Similar, but different, relations are expected in other MOND
formulations. For example, the whole class of modified-inertia
formulation of MOND (Milgrom 1994) predict the following virial
relation for disk galaxies: Define a normalized, effective surface
density of the disk, $\Sigma_*(r)$, such that the Newtonian
acceleration is given by\footnote{$\Sigma_*$ is a sort of
``spherical equivalent'' surface density: If $\rho(r)$ is the
spherical density distribution that gives the same Newtonian
acceleration as the disk, then $\Sigma_*=2r\rho/M$.}
 \beq g_N(r)={V^2(r)\over r}={MG\over r^{2}}\int_0^r2\pi r'\Sigma_*(r')dr', \eeqno{lireq}
and define now the rms rotational velocity weighted with
$\Sigma_*(r)$ instead of $\Sigma(r)$. Then the relation
 \beq \langle V^2\rangle_*^2={4\over 9}MG\az \eeqno{jliy}
is predicted by all modified inertia formulations of MOND for
galaxies in the deep-MOND regime. This can be derived
straightforwardly from the fact that in such theories we have for
circular orbits in the deep-MOND limit $V^4(r)/r^2\az=-d\phi/dr$,
where $\phi(r)$ is the Newtonian potential in the disk.
\par
With accurate enough data we can directly test these two classes of
theories without having to calculate rotational speeds, which in the
case of the nonlinear Bekenstein \& Milgrom (1984) formulation is
rather demanding (all the quantities appearing in the relations are
directly observable).

\par
More generally,  scale invariance implies that for very low
acceleration galaxies we have the following scaling property of the
rotation curve. If a disk galaxy with a low surface density $\S(r)$
has a rotation curve $V(r)$, then a disk with surface density
$\a\S(\b r)$ has a rotation curve $\a^{1/4}\b^{-1/2}V(\b r)$
(instead of the Newtonian scaling $\a^{1/2}\b^{-1/2}V(\b r)$)
provided both disks  are in the deep-MOND regime.
\subsection{The external-field effect}

Sometimes one wants to describe a small, self gravitating system
that is itself falling in a field of a mass external to it. A case
in point is a globular cluster, or a dwarf spheroidal galaxy,
falling in the field of a mother galaxy. When the extent of the
system is small compared with the scale over which the external
field varies, and the intrinsic time scales are much shorter than
the characteristic fall time, we may describe the small system,
approximately, as embedded in a constant (in space and time)
acceleration field of the external body. The non-linearity of MOND
leads, generically, to palpable effects of the external field on the
internal motions in the system. This is known as the external-field
effect (EFE) in MOND (Milgrom 1983). The system is controlled by the
same MOND theory with boundary conditions of a constant acceleration
field at infinity, in the case of modified gravity, or a system of
equations transformed to a constantly accelerated frame, otherwise.
The details do not concern us here; all we reckon with is the fact
that in some way a new ``constant'' is added to the MOND theory:
$\vgz$, the vector of external acceleration field. The equations of
motion (\ref{ii}) can now be written symbolically:
 \beq\mathcal{F}_k[m_i,G,\az,\vgz,\vr_i(t)]=0,~~~~k=1,2,3,...~~.\eeqno{iimod}
 Note that rotational symmetry is now broken. While $\vgz$ is a
 constant of the effective theory thus obtained, unlike
 $\az$, it does transform under scaling. The reason is that scale
 invariance holds only if we apply scaling to the whole system,
 including the masses that give rise to $\vgz$, and this will scale
 $\vgz\rar\l^{-1}\vgz$. Let us again work in units in which
 $G=\az=1$ are dimensionless; so the equations of motion are written\footnote{I use
$\mathcal{F}_k$ in all the equations that follow even though their
meaning changes somewhat from equation to equation.}
 \beq\mathcal{F}_k[m_i,\gz,\vr_i(t)]=0,~~~~k=1,2,3,...~~,\eeqno{iimodi}
where I also suppressed the dependence on the direction of $\vgz$.
Now apply a change in the time units by a factor $\xi$, and scaling
by a factor $\l=\xi^{-1}$. Scale invariance then tells us that the
equations of motion
 \beq\mathcal{F}_k[\xi^4m_i,\xi\gz,\xi\vr_i(t)]=0,~~~~k=1,2,3,...~~,\eeqno{iimodii}
are also satisfied. Define the quantities
$\tilde\vr_i(t)=\vr_i(t)/\gz,~~\tilde m_i=m_i/\gz^4$, and we have
that
 \beq\mathcal{F}_k[\tilde m_i,\xi\gz,\tilde\vr_i(t)]=0,~~~~k=1,2,3,...~~.\eeqno{iimodiv}
Namely, if $\tilde\vr_i(t)$ are a solution for a system with given
$\tilde m_i$ and $\gz$, then it is also a solution for the value
$\xi\gz$, with arbitrary $\xi$, and the same $\tilde m_i$. In other
words, in the equations expressed in terms of $\tilde m_i$ and
$\tilde\vr_i(t)$,  $\gz$ disappears. Scale invariance thus tells us
that the equations of motion can be cast in the form
 \beq\mathcal{F}_k[\tilde m_i,\tilde\vr_i(t)]=0,~~~~k=1,2,3,...~~\eeqno{iimodv}
(the dependence on the direction of $\vgz$ is still there). In our
units $\vr_i$, as well as $\tilde\vr_i$, have dimensions of
time$^2$, while $m_i$ and $\tilde m_i$ have dimensions of time$^4$.
The second time derivatives of $\tilde\vr_i$ (accelerations
normalized by $\gz$) are thus dimensionless as are quantities of the
construction $\tilde m/\tilde r^2$. We can thus write schematically
an expression that describes the general scaling of accelerations in
the system with masses, size, and $\gz$:
  \beq \tilde a=f(\tilde m/\tilde r^2).  \eeqno{accat}
Putting back the dimensioned constants $G$ and $\az$ we can write
this as
 \beq a=\gz f\left({mG\az\over\gz^2r^2}\right). \eeqno{hutre}
If it where not for scale invariance, $a/\gz$ could also depend on
another dimensionless variable such as $\gz/\az$, or other
equivalent ones, but the argument of $f$ in eq.(\ref{hutre}) is the
only combination that is both dimensionless and scale invariant (as
is $a/\gz$). Note that eq.(\ref{hutre}), while quite restrictive,
still leaves much latitude for the effects of the external field, as
$f$ is not constrained, and can vary from theory to theory. For
example, $f(x)\propto x$ describes the EFE in the original
discussion of the effect in Milgrom (1983), and in the formulation
of Bekenstein and Milgrom (1984). It gives a quasi-Newtonian
behavior: $a\propto MG_e/r^2$, with an effective $G_e=G\az/\gz$.
This is only the general scaling law; details-such as the angular
dependence around the direction of $\vgz$--differ in the two
theories. Scaling alone does not even imply an EFE, as $f(x)\propto
x^{1/2}$ corresponds to a theory with no EFE, and gives the standard
scaling for the deep-MOND limit in an isolated system. Note that the
meaning of $\vgz$ as an external field has not played a role in the
considerations above. All that was assumed is that the system is
characterized by an additional constant with units and scaling
dimensions of an acceleration.

\section{The deep-MOND limit as a zero-rest-mass limit}
\label{sect:zromass} It is well known that the appearance of finite
rest masses in a theory is an obstacle to scale and conformal
invariance. In the case of MOND the appearance of the new constant
$\az$, is potentially a further hinderance, as it scales as an
inverse mass. The reason is briefly this: $MG$ has dimensions of
$V^2r$; so if all masses were dynamical (as opposed to rest masses,
which are constants of the theory), they would transform like length
under scaling, instead of being considered untouchable constants:
dimensional balance would have then insured that masses do not
obstruct the symmetry. When considering conformal invariance under a
change in metric $g_{\m\n}\rar \kappa^2(x)g_{\m\n}$, the
corresponding requirement is that rest masses transform as $M\rar
\kappa^{-1}(x)M$, as a necessary condition for the symmetry to hold
(see, e.g., Fulton et al. 1962, Bekenstein and Meisels 1980). But
since rest masses (and $\az$) are generally considered ``constants
of nature'', one is loath to have them transform in these ways, in
which case one has to forgo the symmetry in theories with rest
masses. However, since in the deep-MOND limit we only have products
$m_i\az$ appearing, they do not obstruct the symmetry. As I discuss
below, there are other, ``dynamical'', masses that appear in MOND
and that are not multiplied by $\az$. These, however, automatically
transform appropriately for the symmetry to remain unimpaired.
\par
The deep-MOND limit may also be viewed as a limit of zero rest
masses: Since in this limit rest masses appear only as
$m_i\az$--constants that control the dynamics of the system and stay
fixed when taking the formal limit $\az\rar\infty$--it follows that
the same formal limit corresponds to $m_i\rar 0$. This does not
mean, of course, that the effects of gravity vanish in the limit.
They are taken over by the ``phantom'' masses: the quantities with
the dimensions of mass that govern the motions, and that would be
attributed to dark matter by a dark matter advocate. In the
deep-MOND limit, the rest masses become negligible compared to the
``phantom masses'', hence the formal limit of zero rest masses.
\par
In nonrelativistic, modified gravity theories the ``phantom'' masses
are easily envisaged. Let $\phi$ be the MOND potential produced by a
rest mass distribution $\rho(\vr)$. The same potential will be
produced in Newtonian gravity by a ``dynamical'' density
 \beq \rho_d\equiv(4\pi G)^{-1}\Delta\phi,  \eeqno{puis}
and the difference $\rho_*=\rho_d-\rho$, the so called ``phantom''
density, would be attributed to ``dark matter'' by those who prefer
it\footnote{There is no guarantee that $\rho_*$ satisfies the
standard positivity conditions; in fact, there are examples to the
contrary.}. In theories that cannot be described as ``modified
gravity'' there is no unique distribution of ``phantom matter'' in a
given system, but dynamics can still be described as being
effectively controlled by quantities with the dimensions of mass.
All such quantities have to scale as $GM_*\sim (MG\az)^{1/2}\ell$,
where $M$ is a rest mass parameter, and $\ell$ is some
characteristic length of the system, or of the motion, such as
system size or orbital size. Given that rest masses can appear only
in the combination $MG\az$, this is the only way to construct a mass
from the quantities available. Such mass quantities have two
important properties: they remain finite in the deep-MOND limit ,
and they scale as length, so they do not obstruct scale invariance.
\par
As an example, consider the MOND acceleration, say on a circular
orbit, at a distance $r$ from a point mass, $M$, which is given by
 \beq a={(M+M_*)G\over r^2}, \eeqno{monac}
where $M_*$ is just a parametrization of the departure from the
Newtonian acceleration given by the first term. For small radii
$M_*$ goes to zero; for radii $r\gg r_t$--with $r_t=(MG/\az)^{1/2}$
being the transition radius--$M_*$ dominates and approaches the
value $M_*\approx Mr/r_t\gg M$, which indeed scales as described
above. Equation (\ref{monac}) is not scale invariant for finite $M$,
but becomes so in the deep-MOND limit, where $M/M*\rar 0$.
\par
To summarize, starting with a MOND theory that is characterized by
the constants $Gm_i$ and $\az$, express these parameters by
$q_i\equiv m_i/M$, $S\equiv MG\az$, and $M$ ($M=\sum m_i$, so $\sum
q_i=1$). The MOND limit is gotten by substituting $m_i=0$,
everywhere, with $q_i$ and $S$ fixed.

\section{Possible cosmological connections}
\label{sect:cosmo} There is a well known numerical proximity between
$\az$ and cosmological acceleration scales. Observations suggest two
such potentially relevant scales, one associated with the expansion
rate $a_H\equiv cH_0$, where $H_0$ is the Hubble constant, and
another associated with the measured ``cosmological constant'',
$\haz\equiv c(\L/3)^{1/2}$ ($\L$ here having units of $t^{-2}$). It
so happens that these two are very near in value, in the present
cosmological epoch (the ``cosmological coincidence''), as expressed
by the fact that $\Omega_\L\equiv \L/3H_0^2$ is today near 1. Since
$\az$ is a very prominent, universal, measured signature of galactic
dynamics, it should be viewed as another mysterious coincidence
(e.g., Milgrom 1983, 1989) that
 \beq\baz\equiv 2\pi\az\approx a_H\approx \haz. \eeqno{kipt}
This may bespeak a physical connection between MOND and the entity
behind $\L$ and/or the expansion rate.
\par
Instead of working with the constant $\az$ we could work with a mass
constant $M_0\equiv c^4/G\az$. From relation (\ref{kipt}), we have
today $M_0\approx 2\pi M_U$, where $M_U\equiv
c^3G^{-1}(\L/3)^{-1/2}\approx c^3G^{-1}H_0^{-1}$ is the
characteristic energy in the universe within the horizon today. Our
findings above are then tantamount to having in the deep-MOND limit
all masses appear in the equations of motion only in the ratios
$m_i/M_0$. For example, the mass-velocity relation predicted by MOND
eq.(\ref{tf}) takes the form \beq (V_{\infty}/c)^4=M/M_0. \eeqno{mv}
\par
We need a dynamical theory that connects MOND with cosmology to
deduce, among other things, which of the cosmological acceleration
scales, if any of the two, is to be identified with $\az$. In fact,
$\az$ could depend in a more complicated way on the cosmological
state. For example, we can write heuristically
 \beq \baz=a_H \mathcal{A}(\haz/a_H). \eeqno{hious}
Pinpointing such a dependence would have great consequences
regarding the cosmological time variations of $\az$, with obvious
ramifications for both cosmology and galaxy dynamics, as I discussed
elsewhere. Examples of possible cosmological variations of $\az$ in
specific, relativistic MOND formulations was discussed by Sanders
(2005), and by Bekenstein and Sagi (2008). The potential for
constraining such variations observationally was discussed recently
in Milgrom (2008), and in Limbach et al. (2008). One obvious
possibility that has been considered (and also used in structure
formation simulations) is that $\az$ is proportional to $\haz$, and
that the ``dark energy'' is a veritable constant. In this case,
$\az$ would not vary with cosmic time. Other possibilities exist, of
course; for example, the interesting one I discuss below.

 \par
The following facts might, together, prove very relevant to the
MOND-cosmology connection:

1. If the ``dark energy'' is a ``cosmological constant'', it will
become increasingly dominant (over matter) in the future, with
$\Omega_\L\rar 1$. The space-time we live in will then approach an
exact de Sitter geometry (characterized by $\Omega_\L=1$) in the
cosmological future.

2. The symmetry (isometry) group of such an exact, 4-dimensional, de
Sitter space-time, \dSST, is isomorphic (equivalent) to the group of
conformal transformations in 3-dimensional Euclidean space\footnote{
Space then has to be compactified by adding the point at infinity.};
it is also the conformal group acting on the 3-dimensional Euclidean
sphere. The two groups are ten dimensional. In the former
group--analogous (but not equivalent) to the \poin~ group in a flat
space-time--the transformations can be represented by the ten
rotations about the origin in the 5-dimensional, flat, Minkowski
space in which a \dSST~ can be embedded. They can be described as
corresponding, in the limit of an infinite radius of the \dSST~ (or
$\L=0$ with no matter), to three space rotations, three space
translations, one time translation, and three Lorentz boosts. In the
3-dimensional Euclidean space, time clearly disappears, and the
equivalent, conformal group is generated by the three (space)
rotations, three translations, one dilatation, and three so-called
proper conformal transformations (as it were, the four operations
involving time: time translations and boosts, metamorphose into
space dilations and proper conformal transformations in space).

3. The boundary of the 4-dimensional \dSST~ is made of two
3-dimensional Euclidean spheres, one in the infinite past one in the
future, and so the isometry of the \dSST~ space-time can be said to
be equivalent to the conformal symmetry group on its boundary.

In addition, in connection with MOND we know that:

4. MOND is purportedly linked with cosmology, as pointed to by
relation (\ref{kipt}) above, by which the value of $\az$ is related
to the geometry of the universe, in particular to the radius of the
asymptotic de Sitter space-time.

5. As mentioned above (details in Milgrom 1997), the deep-MOND limit
of the gravitational field equation of the Bekenstein \& Milgrom
(1984) formulation has the full symmetry of the conformal group in
3-dimensional Euclidean space. More specifically, the possibility
was raised there that this MOND-limit of the theory could be a
classical limit of some conformal (quantum) field theory (CFT). This
was based on the identification of certain quantities in the theory
as the ``primary fields'', in the jargon of CFT, which come complete
with their correlation functions and anomalous dimensions. The
``correlation functions'', which are related to the energy of a
configuration of $N$ point masses, $m_i$, placed at positions
$\vr_i$, contain all the observable information about gravitational
systems made of massive particles; so, knowing them is tantamount to
solving the problem completely. In this theory for the gravitational
field, time has disappeared, as explained above in connection with
the symmetry, and has to be brought back in (in a still moot manner)
when we consider the equation of motion of masses.
\par
In light of these, one may conjecture that the MOND-cosmology
connection is such that local gravitational physics would take
exactly the deep-MOND form in an exact de Sitter
universe\footnote{This conjecture differs from another, earlier one
by which MOND with its two different limits would apply in an exact
\dSST, with the de Sitter radius, as represented by $\haz$,
dictating the transition acceleration.}. This is based on the
equality of the symmetry groups of \dSST~ and of the MOND limit of
the Bekenstein-Milgrom formulation (point 2 and 5 above), both
groups being $SO(4,1)$. The fact that today we see locally a
departure from the exact MOND-limit physics--i.e., that the
interpolating functions have the form they have, and that $\az$ is
finite and serves as a transition acceleration--stems from the
departure of our actual space-time from exact \dSST~ geometry: The
broken symmetry of our space-time is thus echoed in the broken
symmetry of local physics. This conjecture would then imply that the
deep-MOND limit will be approached by local physics, as our universe
approaches asymptotically an exact \dSST~ geometry in the
future\footnote{When I speak of the deep-MOND limit as it pertains
to present day physics, I mean its application to systems for which
all accelerations are much smaller than the present value of $\az$,
and whose dynamics can thus be deduced by formally taking $\az$ to
infinity with $MG\az$ fixed. Here, I mean that the limit would
actually be approached in the future, asymptotically \dSST~ universe
for all local systems. So, any gravitational system will go deeper
and deeper into the MOND regime. One has then to consider how the
structure of the system responds to the adiabatic changes involved.
That masses also change, so as to keep $m_i/M_0$ fixed in the limit,
is also a possibility to reckon with.}$^,$\footnote{The deep-MOND
limit might then also prevail during past inflationary phases.}. For
a relation such as eq.(\ref{hious}) this would mean that
$\mathcal{A}\rar\infty$ for $a_H/\haz \rar 1$. But, more generally,
local physics would have to be able to sense, somehow, whether the
local space-time neighborhood is nearly pseudo-spherical with
positive curvature (i.e., nearly that of a \dSST), and adjust the
value of $\az$ accordingly. Local physics can sense this using only
local indicators such as the derivative of the Hubble constant. For
example, as one out of many possible relations, $\az\propto
cH/(1+q)$ ($H$ being the expansion rate, and $q=-1-\dot H/H^2$ the
deceleration parameter) would give the deep-MOND limit in periods of
exponential growth of the cosmological scale factor (when $\dot H\ll
H^2$), and $\az\sim cH\propto t^{-1}$ in periods of power-law
growth; whereas $\az\propto \haz/(1+q)$ would also give local
deep-MOND behavior during exponential growth, but constant
$\az\sim\haz$ for power-law growth.
\par

Based on point 1-3 above, a conjecture has been launched of a
correspondence between conformal field theories in 3-D Euclidean
space and gravity in \dSST~(starting, to my knowledge, with
Strominger 2001, and Spradlin et al. 2001; see also Medved 2002, and
references therein)\footnote{This is the dS analog in 4-D, of the
more frequently discussed, and more strongly motivated,
correspondence between conformal field theories in 4-D Minkowski
space-time and quantum gravity in space-times built on 5-D Anti dS
space-time--the so called AdS/CFT, or Maldacena, conjecture.}. If
such a correspondence could be found in connection with MOND, it
might point to a relativistic gravity theory with the symmetries of
\dSST~ that leads to the energy functions encapsuling the deep-MOND
limit, and, more generally, show how a compound MOND theory can
follow in the context of our less ideal cosmological space-time.
\par
Whether the actual correspondence between theories in the two spaces
exist or not, the symmetry connection is there and could be highly
pertinent for MOND.

\par
If indeed, as speculated here, local physics is tending to the
deep-MOND limit in the asymptotically dS future, with
$\az\rar\infty$, and with convergence to 3-D conformal invariance,
this invariance would also apply to the deep-MOND limit {\it today};
i.e., to systems that are today deep in the MOND regime as defined
by today's value of $\az$ (such as low surface brightness galaxies,
and the outskirts of all galaxies). This is because both physical
limits are formally effected by taking $\az\rar\infty$ in the field
equations, and would correspond to the same limiting theory. Also
note that while the symmetry gained in the limit is conformal
invariance in 3-D Euclidean space, in the context of MOND this would
require that the gravitational potential has zero scaling dimension,
which, in turn, implies also the space-time scale invariance of the
equations of motion, our main topic here.
\par
In any event, the relation between MOND and cosmology is presumably
not one-way, with only cosmology affecting local dynamics (e.g., by
dictating a value of $\az$). The equations governing cosmology may
well be also affected by the departure from standard physics that
MOND entails; and this needs to be understood as well.

\subsection{A note on extant relativistic theories}
 In light of what I said above,
I believe that a lasting marriage of MOND with the principles of
relativity will emerge only when we understand the MOND-cosmology
connection. In such an eventuality, $\az$  will be derived from the
theory, and so will the various interpolating functions. It is then
not clear that it is meaningful to ask, at present, what the
deep-MOND limit will be like in relativistic MOND. Nevertheless,
relativistic versions of MOND have been propounded, and one may
wonder how the discussion here bears on such theories. The state of
the art of these efforts is the Tensor-Vector-Scalar (TeVeS) theory
formulated by Bekenstein (2004)--extending previous ideas by Sanders
(1997)--hinging gravity on a vector field in addition to the usual
metric tensor and a scalar field. This theory has also been recast
as a special case of so called Generalized Einstein Aether theories
by Zlosnik et al. (2007). Such theories introduce the limiting
acceleration and the interpolating function by hand. Nevertheless,
they are most valuable as intermediate steps, perhaps pointing the
way to more fundamental MOND theories. Inasmuch as these theories
have a nonrelativistic limit that corresponds to the modified
Poisson equation discussed above, they enjoy its symmetries in the
deep MOND limit of their {\it nonrelativistic} limit. However, As I
have stressed several times in the past, because of relation
(\ref{kipt}), a system that is both in the relativistic strong field
limit, and is deep in the MOND regime, must have a size much larger
then the Hubble radius. Cosmology itself is only at the boundary in
the MOND sense, with typical accelerations of order $\az$. It is
thus difficult to give a meaning to a relativistic deep-MOND limit,
let alone have any observational guidance regarding such systems.
Our topic here is the very deep MOND limit, and there is little of
it that we can apply in a sensible way to relativistic theories per
se.
\par
I just note, in this connection, that the appearance of the speed of
light, $c$, in a MOND based theory does not, in itself, spoil
space-time scale invariance, since the value of $c$ does not change
under simultaneous scaling of the units of length-time. A
relativistic theory could, however, not be invariant to scaling, and
still have an invariant nonrelativistic limit,  with the symmetry
breaking vanishing in the limit $c^{-1}=0$.

\section{Discussion}
\label{sect:disc}
I note, in passing, that
various hints finger space-time conformal maps
(not the Euclidean space transformations discussed above) as being relevant to the various
MOND theories. One hint stems from the discussion here; in
particular, the fact that scaling is one of the conformal maps of
space-time. Also, in Milgrom (2005) I suggested that the symmetries
underlying MOND may have to do with transformations that map
inertial world lines in dS space-time into constant
accelerations ones (another suggestion of a MOND-dS connection). As it turns out, these are the non-rigid
conformal transformations\footnote{It has long been realized that
conformal transformations map inertial world lines in Minkowski
space-time to accelerated ones, as discussed, for example in Fulton
et al. 1962, and in references therein.}. Such conformal maps may
thus play a role in connection with MOND similar to that of Lorentz
transformations in connection with relativity.
\par
 I finally note that the scaling symmetry of the
deep-MOND limit is not forced by dimensional consideration alone, as
is the case for Newtonian gravity. Starting with the latter, recall
that the equations describing Newtonian, purely gravitational
systems are invariant to $(t,\vr)\rar(\l t,\l^{2/3}\vr)$. This is
because $m_iG$, the only dimensioned constants that appear, retain
their values under the corresponding change of units\footnote{This
scaling property remains valid for nonrelativistic electrostatics,
since the action describing a system of masses and charges can be
written such that the only dimensioned constants appearing are of
the type $mG$, and $e^2/m$, which is of the same dimensions.}.
General Relativity does not enjoy the same symmetry because $c\rar
\l^{-1/3}c$; but its limit $c^{-1}=0$ does.
\par
In MOND, we have two types of dimensioned constants, $m_iG$ and
$\az$. So, in the limit $\az\rar\infty$, we could have had
invariance to any scaling of the form $(t,\vr)\rar(\l t,\l^\b\vr)$.
This would have lead to $\az$ and masses appearing in the
combination $MG\az^{1+\a}$  [$\a=4(\b-1)/(2-\b)$], since these are
invariant in value under the change of units that corresponds to the
above scaling. Different phenomenologies would have ensued. For
example the asymptotic rotational speed around a mass $M$ would be
 \beq V(r)\rar (MG)^{{2-\b\over 4\b}}\az^{{3\b-2\over
 4\b}}r^{{\b-1\over \b}}=(MG\az^{1+\a}r^\a)^{{1\over 4+2\a}}. \eeqno{altera}
For Newtonian gravity $\b=2/3$ ($\a=-1$), which gives
straightforwardly Kepler's third law, and the dependence of the
Kepler constant on the central mass. The requirement of
asymptotically flat rotation curves--the founding axiom of
MOND--dictates $\a=0,~\b=1$.

\section*{Acknowledgements}

 This research was supported by a
center of excellence grant from the Israel Science Foundation. I
thank Jacob Bekenstein for insightful comments.

\bibliographystyle{elsart-harv}
\bibliographystyle{aa}

\begin{thebibliography}{36}
 \expandafter\ifx\csname natexlab\endcsname\relax\def\natexlab#1{#1}\fi




\bibitem[Bekenstein (2004)]{bek04} Bekenstein, J.D. 2004, Phys.Rev. D70  083509
\bibitem[Bekenstein (2006)]{bek06} Bekenstein, J.D. 2006, Contemp. Phys., 47, 387
\bibitem[Bekenstein \& Meisels (1980)]{bmeis80}
Bekenstein, J.D. \& Meisels, A. 1980, Phys. Rev. D, 22, 1313
\bibitem[Bekenstein \& Milgrom (1984)]{bm84}
Bekenstein, J. \& Milgrom, M. 1984, ApJ, 286, 7
\bibitem[Bekenstein \& Sagi (2008)]{bs08}Bekenstein, J.D. \& Sagi, E.
2008,  Phys. Rev. D, 77, 103512
\bibitem[Fulton Rohrlich and Witten 1962]{frw62}Fulton, T.,
Rohrlich, F., and Witten, L. 1962, Rev. Mod. Phys. 34 (3), 442
\bibitem[Limbach \& al. 2008]{limbach08}Limbach, C., Psaltis, D., \&
Ozel, F. 2008,  arXiv:0809.2790
\bibitem[Medved 2002]{medved02} Medved, A.J.M. 2002, Class. Quant. Grav. 19,
4511, hep-th/0203191
\bibitem[Milgrom (1983)]{mil83} Milgrom, M. 1983, ApJ, 270, 365
\bibitem[Milgrom (1989)]{mil89} Milgrom, M. 1989, Comm. on Astroph., 13, 215
\bibitem[Milgrom (1994)]{mil94} Milgrom, M. 1994, Ann. Phys., 229, 384
\bibitem[Milgrom (1997)]{mil97} Milgrom, M. 1997, Phys. Rev. E, 56,
1148
\bibitem[Milgrom (2001)]{mil01} Milgrom, M. 2001, Acta Phys. Polon. B, 32, 3613
\bibitem[Milgrom (2005)]{mil05} Milgrom, M. 2005, Proceedings of the XXIst IAP Colloquium
``Mass Profiles and Shapes of Cosmological Structures'', Eds. G.
Mamon, F. Combes, C. Deffayet, B. Fort, EDP Sciences,
arXiv:astro-ph/0510117
\bibitem[Milgrom (2008)]{mil08} Milgrom, M., 2008, In Proceedings XIX Rencontres de
Blois; arXiv:0801.3133
\bibitem[Sanders (1997)]{sanders97}Sanders, R.H. 1997, ApJ, 480, 492
\bibitem[Sanders (2005)]{sanders05}Sanders, R.H. 2005, MNRAS, 363, 459
\bibitem[Sanders \& McGaugh (2002)]{sm02}Sanders, R.H. \& McGaugh, S.S. 2002, ARAA, 40, 263
\bibitem[Spradlin Strominger \& Volovich 2001]{ssv01} Spradlin, M.,  Strominger, A., \&
Volovich, A. 2001, Les Houches Lectures on De Sitter Space,
arXiv:hep-th/0110007
\bibitem[Strominger (2001)]{strom01}Strominger, A. 2001, J. High
Energy Phys. 110, 34
\bibitem[T.G Zlosnik \& al. 2007]{z07}Zlosnik, T.G., Ferreira, P.G., \&
Starkman, G.D. 2007 Phys. Rev. D75, 044017

\end{thebibliography}

\clearpage
\end{document}